\documentclass[12pt]{article}
\usepackage[english]{babel}\usepackage[latin1]{inputenc}  
\usepackage{epsfig}
\usepackage[usenames]{color}

\title{\Large {\bf ON THE CONNECTION BETWEEN THE LIOUVILLE EQUATION AND THE SCHRÖDINGER EQUATION}}

\author{Edelver Carnovali Jr\footnote{Email address: edelver@gmail.com} and 
Humberto M. França\footnote{Email address: hfranca@if.usp.br}}

\date{{\em Instituto de Física, Universidade de São Paulo, Caixa Postal 66318, 05315-970
São Paulo, Brasil}}

\begin{document}\maketitle

{\bf Abstract}
\bigskip

We derive a classical Schrödinger type equation from the phase space Liouville equation valid
for an arbitrary nonlinear potencial $V(x)$. 
The derivation is based on a Wigner type Fourier transform of the classical phase space 
probability distribution, which depends on a small constant $\alpha$ with dimension
of action, that will be characteristic of the microscopic phenomena. 
In order to obtain the Schrödinger type equation, two requirements are necessary: 1) It is assumed 
that the appropriately defined classical probability 
amplitude $\Psi(x,t)$ can be expanded in a complete set of functions $\Phi _n(x)$ defined 
in the configuration space; 2) the classical phase space distribution $W(x,p,t)$ obeys the Liouville 
equation and is a real function of the position, the momentum and the time. 

We show that for $\alpha$ equal to the Planck´s constant $\hbar$, the evolution equation for the {\em classical} 
probability amplitude $\Psi(x,t)$ is identical to the Schrödinger equation. The wave particle duality principle, 
however, is not introduced in our paper. 
\bigskip



\newpage

\section{Introduction}
\bigskip

In 1932 E. Wigner published an important paper {\cite{wigner,moyal}} where he 
introduced what is called today {\em the Wigner's quasi probability function}, or simply 
{\em Wigner's phase space function}. We shall denote it by $Q(x,p,t)$ and it is connected
with the Schrödinger wave equation solutions $\psi(x,t)$ by the expression:
\begin{equation}\label{wf}
Q(x,p,t)=\frac{1}{\pi\hbar}\int \psi ^*(x-y,t) \psi(x+y,t) e^{-\frac{2ipy}{\hbar}}dy,
\end{equation}
where $\hbar$ is the Planck´s constant. We recall that Wigner´s intention was to obtain a quantum mechanical description of
the classical phase space phenomena.

It was stated later, by Wigner and O'Connel {\cite{wignerprova}}, that the definition (\ref{wf})
is the only one that satisfies a set of conditions that are expected on physical grounds.

Notice, however, that the Planck's constant $\hbar$ enters in the above expression in a combination that was not completely
clarified by Wigner and other authors [1-7]. 
We know that the Planck´s constant $\hbar$ is related to the {\em electromagnetic fluctuation phenomena} characteristic of
Quantum Electrodynamics \cite{dalibard} and Classical Stochastic Electrodynamics [9-12]. However, for simplicity, these electromagnetic 
fluctuations will be not considered in our paper. 

We would like to mention that, if applied to an arbitrary solution $\psi(x,t)$ of the Schrödinger wave equation, the
expression (\ref{wf}) may lead to {\em negative}, or even {\em complex}, phase space probability densities, suggesting that
many solutions of the Schrödinger wave equation cannot be associated with genuine phase space probability
distributions, that is, the relation (\ref{wf}) is problematic. 
Simple examples are the excited states of the harmonic oscillator \cite{franca}. 
Here, following França and Marshall \cite{franca}, we shall consider that these excited states 
are simply useful mathematical functions. We shall clarify this point later.

It is also well known \cite{wigner,woong} that $Q(x,p,t)$ defined in (\ref{wf}) propagates in time according
to the quantum Liouville equation
\begin{equation}\label{aliouville}
\frac{\partial Q}{\partial t}=-\frac{p}{M}\frac{\partial Q}{\partial x}
+\frac{\partial V}{\partial x}\frac{\partial Q}{\partial p}+
\frac{
\left(
\frac{\hbar}{2i}
\right)^2
}{3!}
\frac{\partial ^3 V}{\partial x^3}\frac{\partial ^3 Q}{\partial p^3}
+
\frac{
\left(
\frac{\hbar}{2i}
\right)^4
}{5!}
\frac{\partial ^5 V}{\partial x^5}\frac{\partial ^5 Q}{\partial p^5}+...,
\end{equation}
provided that $\psi(x,t)$ satisfies the Schrödinger wave equation. Here  
$F(x)=-\partial V/\partial x$ is the arbitrary nonlinear force acting on the particle with mass $M$. 

The equation (\ref{aliouville}) shows 
that the definition (\ref{wf}), and the Schrödinger equation, do not lead to the classical 
Liouville equation. We recall that, according to the correspondence principle, one should 
necessarily obtain the Liouville equation if the mass $M$ is large enough. Therefore, the relationship between the
equation (\ref{wf}), the Schrödinger equation and the Liouville equation is also problematic.  

In our paper, starting with the classical Liouville equation, we shall see the mathematical and the physical conditions 
to connect it with an equation which is formally identical to the Schrödinger equation for a {\em classical} probability 
amplitude $\Psi(x,t)$ defined in the configuration space. 

By means of the theorem established within the section 2,
we show that the connection between the Liouville equation and the Schrödinger type equation is valid for an arbitrary 
potential $V(x)$. This connection was presented in reference  
\cite{manfredi}, restricted to the case where the potential $V(x)$ is quadratic in the variable $x$. 

Our presentation is organized as follows. We give, within section 2, the mathematical background
necessary to connect the classical phase space probability distribution to the classical 
configuration space probability amplitude $\Psi(x,t)$. 
For clarity sake, we shall denote $\psi(x,t)$ a solution of the Schrödinger wave equation and $\Psi(x,t)$ the classical 
probability amplitude.
The direct relation between the Liouville equation for an arbitrary potential $V(x)$,
and the classical Schrödinger type equation is presented within section 3. A brief discussion and our conclusions are 
presented in the final section.

\section{Mathematical background}
\bigskip

Our starting point is the Liouville equation for the probability distribution in phase 
space, denoted by $W(x,p,t)$. The Liouville equation is such that 
\begin{equation}\label{liouvillea}
\frac{\partial W}{\partial t}+\frac{p}{M}\frac{\partial W}{\partial x}+
F(x)\frac{\partial W}{\partial p}=0.
\end{equation}

The nonlinear force acting on the particle with mass $M$ will be denoted by $F(x)=-\partial V(x)/\partial x$ and $V(x)$ 
is an arbitrary continuous function. 
The solutions of the equation (\ref{liouvillea}) are obtained in conjunction with the solutions 
of the Newton equation
\begin{equation}\label{newton}
M\frac{d^2x}{dt^2}=F(x),
\end{equation}
and the classical positive distribution function, $W_0(x_0,p_0)$, associated with the initial conditions $x_0$ and
$p_0$, namely
\begin{equation}\label{wzero}
W(x,p,t)=\int dx_0 \int dp_0 W_0(x_0,p_0)\delta (x-x_c(t))\delta (p-p_c(t)).
\end{equation}

Here $x_c(t)$ is the classical trajectory and $p_c(t)=m \dot{x}_c(t)$. Notice that $x_c(t)$, and $p_c(t)$, 
depend on $x_0$ and $p_0$. 
Therefore, $W(x,p,t)$ is always real and positive, and the classical probability amplitude 
$\Psi(x,t)$ is such that
\begin{equation}\label{psi2}
|\Psi(x,t)|^2=\int^{\infty}_{-\infty}dpW(x,p,t) .
\end{equation}
The above definition is not enough to determine the complex amplitude $\Psi(x,t)$. To achieve
this goal we shall use the Liouville equation (\ref{liouvillea}).

The Liouville equation for $W(x,p,t)$ has a close relation with a classical Schrödinger type 
equation
for $\Psi(x,t)$, as we shall see in the next section. 
In order to clearly explain this relationship we shall establish a useful theorem 
based on a convenient Fourier transform of $W(x,p,t)$ similar to that used in equation (\ref{wf}).
\bigskip

\underline{{\em Theorem}}: 
\bigskip

Consider the Fourier transform defined by
\begin{equation}\label{twa}
T[W](x,y,t)=\int _{-\infty}^{\infty} W(x,p,t)e^{\frac{2ipy}{\alpha}}dp,
\end{equation}
where $\alpha$ is a {\em small} constant with dimension of action.
The mathematical definition (\ref{twa}) has a physical meaning only when $y\approx 0$ because in this case, the equation 
(\ref{twa}) becomes identical to (\ref{psi2}) and gives us the probability distribution in the configuration space.
Mathematically, we can write $T[W](x,y=0,t)=|\Psi(x,t)|^2$. We shall show that $T[W](x,y,t)$
can always be expressed in the form
\begin{equation}\label{psipsi}
T[W](x,y,t)=\Psi ^*(x-y,t)\Psi(x+y,t),
\end{equation}
provided that $W(x,p,t)$ is a {\em real function} of its variables.
\bigskip

\underline{{\em Proof}}:
\bigskip

Consider any set of functions $\Phi _n(x)$, that satisfy the completeness relation
\begin{equation}\label{completeza}
\sum _{n}\Phi^*_n(x)\Phi _n(y)=\delta(x-y),
\end{equation}
and are orthogonal to each other. An example of these functions is the set
\begin{equation}\label{funcaophi}
\Phi _n(x)=\frac{1}{\sqrt{2b}}e^{\frac{in\pi x}{b}} ,
\end{equation}
where $n=0,\pm 1,\pm 2,\pm 3,...$ and $b$ is a positive constant ($-b < x < b$). Other sets
of functions $\Phi _n(x)$ can be used without loss of generality \cite{moyal,franca}.

Therefore, any complex probability amplitude $\Psi(x,t)$ can be expressed as 
\begin{equation}\label{7a}
\Psi(x,t)=\sum _n a_n (t)\Phi _n(x) ,
\end{equation}
where $a_n(t)$ are the coefficients of the expansion. Different sets of coefficients $a_n(t)$ will give different
probability amplitudes $\Psi(x,t)$. One can show, from the orthogonality properties of the functions $\Phi _n(x)$,
that 
\begin{equation}\label{aene}
a_n(0)=\int dx \Phi^*_n(x)\Psi(x,0) .
\end{equation}

Let $W_{mn}(x,p)$ be phase-space functions defined by \cite{moyal,franca}
\begin{equation}\label{wmn}
W_{mn}(x,p)=\frac{1}{\pi\alpha}\int dy \Phi^*_m(x-y)\Phi _n(x+y)e^{-\frac{2ipy}{\alpha}},
\end{equation}
where $\alpha$ is the same constant introduced in (\ref{twa}).
This is a convenient mathematical definition because the functions $W_{mn}(x,p)$ have 
useful properties as we shall see in what follows. Notice that the functions $W_{mn}(x,p)$
can be {\em positive, negative or even complex functions} \cite{moyal,franca}.

The functions $W_{mn}(x,p)$ constitute a complete and orthogonal set of functions in 
phase space.
The completeness can be verified as follows:
$$
\sum _{m}\sum _{n}W^*_{mn}(x,p)W_{mn}(y,q)=
$$
$$
=\frac{1}{(\pi\alpha)^2}\int d\xi\int d\eta e^{-\frac{2i}{\alpha}(q\eta-p\xi)}
\sum _{m}\Phi _m(x-\xi)\Phi^*_m(y-\eta)
\sum _{n}\Phi _n(y+\eta)\Phi^*_n(x+\xi)
$$
\begin{equation}\label{completa}
=\frac{1}{\pi\alpha}\delta(x-y)\delta(q-p).
\end{equation}
In the last equality the fact that functions $\Phi _n(x)$ form a complete set was used,
and we also used the Fourier integral representation for the Dirac's delta 
function.

A verification of the orthogonality requirement can be obtained in an 
analogous way and leads us to the following relation \cite{moyal,franca}:
\begin{equation}\label{ortogonal}
\int dx\int dp W^*_{mn}(x,p)W_{rs}(x,p)=\frac{1}{\pi\alpha}\delta _{mr}\delta _{ns}.
\end{equation}

Therefore, it is always possible to put  the classical phase-space distribution $W(x,p,t)$ in the form
\begin{equation}\label{wexpand}
W(x,p,t)=\sum _{m}\sum _{n}C_{mn}(t)W_{mn}(x,p) ,
\end{equation}
where the coefficients $C_{mn}(t)$ are only functions of the time $t$ \cite{franca}.

By substituting the explicit form of $W_{mn}(x,p)$ in the last equation, we obtain the formal expression
\begin{equation}\label{wmnb}
W(x,p,t)=\frac{1}{\pi\alpha}\int dy \sum _{m}\sum _{n}
C_{mn}(t)\Phi^*_m(x-y)\Phi _n(x+y)e^{-\frac{2ipy}{\alpha}} .
\end{equation}

Since $W(x,p,t)$ is a {\em real} function we have 
\begin{equation}\label{wreal}
W(x,p,t)=W^*(x,p,t) .
\end{equation}

Using the relation (\ref{wreal}) we get the following equality:
$$
\int dy \sum _{m}\sum _{n}
C_{mn}(t)\Phi^*_m(x-y)\Phi _n(x+y)e^{-\frac{2ipy}{\alpha}}=
$$
\begin{equation}\label{igualdade}
=\int dy \sum _{m}\sum _{n}
C^*_{mn}(t)\Phi _m(x-y)\Phi^*_n(x+y)e^{\frac{2ipy}{\alpha}} .
\end{equation}
Considering the change of variables $y=-\xi$, and exchanging the dummy indices $m$ and $n$,
we can write (\ref{igualdade}) in the form
$$
\int dy \sum _{m}\sum _{n}
C_{mn}(t)\Phi^*_m(x-y)\Phi _n(x+y)e^{-\frac{2ipy}{\alpha}}=
$$
\begin{equation}\label{eq15}
=\int d\xi \sum _{m}\sum _{n}
C^*_{nm}(t)\Phi^*_m(x-\xi)\Phi _n(x+\xi)e^{-\frac{2ip\xi}{\alpha}}.
\end{equation}

From this equation we conclude that
\begin{equation}\label{cautadj}
C_{mn}(t)=C^*_{nm}(t) .
\end{equation}

Therefore, the elements $C_{mn}(t)$ can be written in the following form:
\begin{equation}\label{coeficientes}
C_{mn}(t)=a^*_m(t)a_n(t) ,
\end{equation}
where the functions $a_n(t)$ are only functions of the time $t$ \cite{franca}. In order to make the connection with 
the probability amplitude $\Psi(x,t)$ we shall assume that the functions $a_n(t)$ are the same functions introduced above 
in the equation (\ref{7a}).

Using (\ref{coeficientes}), (\ref{wmnb}) and (\ref{twa}) we obtain
$$
\int dp W(x,p,t)e^{\frac{2ipy}{\alpha}}=
$$
$$
=\frac{1}{\pi\alpha}\int dy'
\left[
\sum _{m}a_{m}^{*}(t)\Phi _{m}^{*}(x-y')
\right]
\left[
\sum _{n}a_{n}(t)\Phi _{n}(x+y')
\right]
\int dp e^{\frac{2ip}{\alpha}(y-y')}
$$
\begin{equation}\label{fouriertransf}
=
\left[
\sum _{m}a_{m}^{*}(t)\Phi _{m}^{*}(x-y)
\right]
\left[
\sum _{n}a_{n}(t)\Phi _{n}(x+y)
\right]
\end{equation}

So, according to the definition (\ref{twa}) and the expression (\ref{7a}) for the
classical probability amplitude, we get
\begin{equation}\label{psipsi2}
T[W](x,y,t) \equiv \int^{\infty}_{-\infty} dp W(x,p,t)e^{\frac{2ipy}{\alpha}}=
\Psi^*(x-y,t)\Psi(x+y,t) ,
\end{equation}
thus completing the demonstration of the theorem.

\section{Considerations concerning the Liouville equation and its connection with the 
Schrödinger equation}
\bigskip

Phase space distribution functions, as those defined in (\ref{wmn}),
provide a framework for a reformulation of non-relativistic quantum mechanics (QM)
in terms of classical concepts \cite{wigner,moyal,franca}.

Moreover, it is a widespread belief that the connection between the 
Liouville equation and the Schrödinger type equation is only possible for quadratic potentials in the $x$ variable
($V(x,t)=a(t)x^2+b(t)x+c(t)$, where $a(t)$, $b(t)$ and $c(t)$ are arbitrary functions of
time \cite{manfredi}). We shall explain, in this section, that the potential energy $V(x)$
can be an arbitrary function of $x$. In order to achieve this goal we shall use a procedure
discussed before by L. S. Olavo \cite{olavo} and K. Dechoum, H. M. França and C. P. Malta
\cite{francab}. We shall see that the theorem established in the previous section is a proof
of a working hypothesis (see (\ref{psipsi})) used in these and many other works \cite{olavo,francab}. The first work 
quoted in the reference \cite{francab} treats the Stern-Gerlach phenomenon. 

As before, $W(x,p,t)$ is the classical phase space probability density associated
with some physical system (see the equations (\ref{liouvillea}) and (\ref{newton})). We 
shall assume that $W(x,p,t)$ is normalized so that
\begin{equation}\label{norma}
\int dx\int dp W(x,p,t)=1 .
\end{equation}

The classical configuration space probability density will be denoted by $P(x,t)$ given by
\begin{equation}\label{probconf}
P(x,t)=\int dp W(x,p,t)\equiv |\Psi(x,t)|^2 .
\end{equation}
where $\Psi(x,t)$ is the classical probability amplitude (see (\ref{psi2})).

We are interested in obtaining a dynamical differential equation for the classical 
probability amplitude $\Psi(x,t)$ based on the fact that $W(x,p,t)$ obeys the classical 
Liouville equation (\ref{liouvillea}).

To obtain this equation, we shall use the Fourier transform
\begin{equation}\label{wtiu1}
T[W](x,y,t)\equiv\int^{\infty}_{-\infty} W(x,p,t)e^{\frac{2ipy}{\alpha}}dp ,
\end{equation}
defined previously in the equation (\ref{twa}).

We recall that the Fourier transform $T[W](x,y,t)$ is a complex function which has
a physical meaning only in the limit $|y|\rightarrow 0$, that is,
\begin{equation}\label{qx}
T[W](x,y=0,t)=|\Psi(x,t)|^2 .
\end{equation}

Therefore, we shall consider the definition (\ref{wtiu1}) only for very small values of $y$.
Our goal is to obtain the differential equation for $\Psi(x,t)$, from the 
differential equation for $T[W](x,y,t)$, valid when $|y|$ is very small.

Our first step is to consider the equation
\begin{equation}\label{dwtiudt1}
\frac{\partial}{\partial t}T[W](x,y,t)=\int^{\infty}_{-\infty} \frac{\partial W}{\partial t}
e^{\frac{2ipy}{\alpha}}dp .
\end{equation}

Using the Liouville equation (\ref{liouvillea}) we obtain 
\begin{equation}\label{dwtiudt2}
\frac{\partial T[W]}{\partial t}=-\int^{\infty}_{-\infty}
\left[
\frac{p}{M}\frac{\partial W}{\partial x}+F(x)\frac{\partial W}{\partial p}
\right]
e^{\frac{2ipy}{\alpha}}dp .
\end{equation}
One can show that the integral
$$
I=-\int^{\infty}_{-\infty}F(x)\frac{\partial W}{\partial p}e^{\frac{2ipy}{\alpha}}dp
$$
is such that
\begin{equation}\label{I}
I=F(x)\frac{2iy}{\alpha}\int^{\infty}_{-\infty} W(x,p,t)e^{\frac{2ipy}{\alpha}}dp .
\end{equation}
Here we have assumed that
\begin{equation}\label{limite1}
\lim _{|p|\rightarrow \infty}W(x,p,t)=0.
\end{equation}
We also see  that
\begin{equation}\label{pop}
pe^{\frac{2ipy}{\alpha}}=-i\frac{\alpha}{2}\frac{\partial}{\partial y}e^{\frac{2ipy}{\alpha}}.
\end{equation}

Substituting (\ref{I}) and (\ref{pop}) in (\ref{dwtiudt2}) we get
\begin{equation}\label{dwtiudt4}
i\alpha\frac{\partial}{\partial t}T[W](x,y,t)=
\left[
\frac{(-i\alpha)^2}{2M}\frac{\partial ^2}{\partial y\partial x}
-2yF(x)
\right]
T[W](x,y,t) .
\end{equation}

Notice that, in accordance with the theorem proved within section 2, 
$T[W](x,y,t)=\Psi^*(x-y,t)\Psi(x+y,t)$.

It is convenient to use new variables, namely $s=x-y$ and $r=x+y$, so that the equation 
(\ref{dwtiudt4}) can be written as
$$
i\alpha \frac{\partial}{\partial t}[\Psi^*(s,t)\Psi(r,t)]=
$$
\begin{equation}\label{dpsipsidt1}
\left[
\frac{(-i\alpha) ^2}{2M}
\left(
\frac{\partial ^2}{\partial r^2}-\frac{\partial ^2}{\partial s^2}
\right)
-(r-s)F
\left(
\frac{r+s}{2}
\right)
\right]\Psi^*(s,t)\Psi(r,t) .
\end{equation}

According to (\ref{qx}), we want an equation for $\Psi(x,t)=[\Psi(r,t)]_{y\rightarrow 0}$, and the 
equivalent equation for $\Psi^*(x,t)=[\Psi^*(s,t)]_{y\rightarrow 0}$. Consequently we shall consider that 
the points $r$ and $s$ are arbitrarily close so that
\begin{equation}\label{tvm}
-(r-s)F
\left(
\frac{r+s}{2}
\right)
= -\int _{s}^{r}F(\xi)d\xi = V(r)-V(s) , 
\end{equation}
in accordance to the mean value theorem. Substituting (\ref{tvm}) into (\ref{dpsipsidt1})
we get the equations
\begin{equation}\label{schrod1}
i \alpha \frac{\partial}{\partial t}\Psi^*(s,t)=
\left[
\frac{-1}{2M}
\left(
-i \alpha \frac{\partial}{\partial s}
\right)^2 -V(s)
\right]\Psi^*(s,t) ,
\end{equation}
and
\begin{equation}\label{schrod2}
i \alpha \frac{\partial}{\partial t}\Psi(r,t)=
\left[
\frac{1}{2M}
\left(
-i \alpha \frac{\partial}{\partial r}
\right)^2 + V(r)
\right]\Psi(r,t) .
\end{equation}

These two equations are, in fact, the same differential equation, namely
\begin{equation}\label{schrod4}
i \alpha \frac{\partial}{\partial t}\Psi(x,t)=
\left[
\frac{1}{2M}
\left(
-i \alpha \frac{\partial}{\partial x}
\right)^2 + V(x)
\right]\Psi(x,t) ,
\end{equation}
which is formally identical to the Schrödinger wave equation. 
Notice, however, that $\Psi(x,t)$ is a {\em classical} probability amplitude, which is conceptually different
from the Schrödinger {\em wave} function $\psi(x,t)$. The classical probability amplitudes $\Psi(x,t)$ are just
mathematical objects \cite{leggett2}. 

The wave-particle duality principle was not used and the small characteristic constant $\alpha$ is, up to this point, 
arbitrary. This constant can be determined from the observation of several phenomena within microscopic domain [10-12]. 
Therefore, we can conclude that $\alpha = \hbar$, where $\hbar$ is the universal Planck´s constant.

It is relevant to observe that if we write
\begin{equation}\label{psiene}
\Psi(x,t)=\Phi _n(x)e^{-i\frac{\epsilon _nt}{\alpha}} ,
\end{equation}
where $\epsilon _n$ are constants with dimension of energy, the equation (\ref{schrod4}) leads to
\begin{equation}\label{schrodautoval} 
\left[
\frac{1}{2M}
\left(
-i \alpha \frac{\partial}{\partial x}
\right)^2 +V(x)
\right]\Phi _n(x)
=\epsilon _n\Phi _n(x) .
\end{equation}

This is a familiar eigenfunction equation . Its solutions $\Phi _n(x)$ constitute
a complete set of orthogonal functions that can be obtained for different potentials $V(x)$. This set
of functions $\Phi _n(x)$ can also be used in our previous equations (\ref{7a}) and (\ref{wmn}) 
(see reference \cite{franca}) showing the consistency between the sections 2 and 3.

We would like to call the reader attention to the fact that the above mathematical treatment can be extended to the
three dimensional case \cite{francab} without additional difficulty.

\section{Discussion}
\bigskip

The connection between the classical Liouville equation for a probability distribution in phase space and the
Schrödinger equation was already established for quadratic potentials \cite{wigner,manfredi}. The extension of the formalism to a 
{\em generic potential} $V(x)$ presented here put all the previous related works \cite{olavo,francab} on 
a solid mathematical ground. 
We get the Schrödinger type equation (\ref{schrod4}), and also Born's statistic interpretation of 
$|\Psi(x,t)|^2$, by considering the Liouville equation, and 
the equations (\ref{twa}), (\ref{dwtiudt4}) and (\ref{dpsipsidt1}) only in the limit $y \rightarrow 0$. In this context
we were able to use the mean value theorem (see equation (\ref{tvm})) to separate (\ref{dpsipsidt1}) in two equivalent equations, 
formally identical to the Schrödinger equation. Since $y$ has to be small, we understand why the inverse transform 
(equation (\ref{wf})), utilized by Wigner, does not guarantee that the phase space function $Q(x,p,t)$ is a true (positive)
probability distribution.

The association of the constant $\alpha$ with the Planck´s constant $\hbar$ 
is natural and inevitable to any reader familiarized with the Schrödinger equation 
used in the microscopic world. As we said above, the explicit 
numerical value of $\alpha$ (or $\hbar$) can be determined by testing the validity of the Schrödinger type equation in 
various phenomena of the microscopic domain \cite{merzbacher}. The connection of the Schrödinger momentum operator
$-i\hbar \partial/\partial x$ with the zero-point vacuum electromagnetic radiation is also interesting. This fact is 
explored by P. W. Milonni in the references \cite{milonni1,milonni2}.

According to A. J. Legget \cite{leggett2}, ``despite the spectacular success of quantum mechanics over the last 80 years
in explaining phenomena observed at atomic and subatomic level, the conceptual status of the theory is still a topic
of lively controversy''. He seems to believe that quantum mechanics is nothing more than a very good mathematical 
``recipe'' \cite{leggett}.

We agree with the above statements. According to the connection between the Liouville equation (\ref{liouvillea}) and 
the Schrödinger type equation (\ref{schrod4}), presented here, one can conclude that the classical probability amplitude
$\Psi(x,t)$ is not associated with a genuine de Broglie wave. As a matter of fact, the existence of de Broglie waves was
questioned recently by Sulcs, Gilbert and Osborne \cite{sulcs} in their analysis  of the famous experiments \cite{arndt} on
the interference of massive particles as the $C_{60}$ molecules (fullerenes).
\bigskip

{\bf Acknowledgment}
\bigskip

We thank Professor Coraci P. Malta for a critical reading of the manuscript and Alencar
J. Faria, José Edmar A. Ribeiro and Gerson Gregório Gomes for valuable comments. 
This work was supported in part by Conselho Nacional de Desenvolvimento Científico e Tecnológico - CNPq - Brazil.

\end{document}